\documentclass[usenatbib]{mn2e}
\usepackage[english]{babel}
\usepackage{amssymb}
\usepackage{amsmath}
\usepackage{multirow}
\usepackage{graphicx} 
\usepackage{float}
\usepackage{nameref}
\usepackage{footmisc}
\usepackage{wrapfig}
\usepackage{wasysym}
\usepackage{times}
\usepackage{textcomp}
\usepackage{epsfig}
\usepackage{longtable}

\title[\src: Magnetar Phase-resolved Analysis and Evidence for a Variable Cyclotron Feature]{The Outburst Decay of the Low Magnetic Field  Magnetar \src: Phase-resolved Analysis and Evidence for a Variable Cyclotron Feature}
\author[Rodr\'iguez Castillo G. A. et al.]{Guillermo A. Rodr\'iguez Castillo,$^{1,2}$ Gian Luca Israel,$^1$ Andrea Tiengo,$^{3,4,5}$ 
  \newauthor  David Salvetti,$^3$ Roberto Turolla,$^{6,7}$ Silvia Zane,$^{7}$ Nanda Rea,$^{8,9}$ Paolo Esposito,$^{3,10}$ 
  \newauthor  Sandro Mereghetti,$^3$ Rosalba Perna,$^{11}$  Luigi Stella,$^1$ Jos\'e A. Pons,$^{12}$  
  \newauthor  Sergio Campana,$^{13}$ Diego G\"{o}tz,$^{14}$ and Sara Motta$^{15}$\\
  $^1$ INAF -- Osservatorio Astronomico di Roma, Via Frascati 33, I-00040, Monte Porzio Catone, Italy\\
  $^2$ Dipartimento di Fisica, Sapienza Universit\`a di Roma, Piazzale Aldo Moro 5, I-00185, Rome, Italy\\
  $^3$ INAF -- IASF Milano, Via E. Bassini 15, I-20133 Milano, Italy\\
  $^{4}$ Istituto Universitario di Studi Superiori, Piazza della Vittoria 15, I-27100 Pavia, Italy\\
  $^{5}$ Istituto Nazionale di Fisica Nucleare, Sezione di Pavia, Via A. Bassi 6, I-27100 Pavia, Italy\\  
  $^{6}$ Dipartimento di Fisica e Astronomia, Universit\`a di Padova, Via Marzolo 8, I-35131, Padova, Italy\\ 
  $^{7}$ Mullard Space Science Laboratory, University College London, Holmbury St. Mary, Dorking, Surrey, RH5 6NT, UK\\
  $^{8}$ Astronomical Institute ``Anton Pannekoek", University of Amsterdam, Postbus 94249, 1090 GE Amsterdam, The Netherlands\\
  $^{9}$ Institute of Space Sciences (CSIC-IEEC), Campus UAB, Carrer de Can Magrans, s/n, 08193 Barcelona, Spain\\
  $^{10}$ Harvard--Smithsonian Center for Astrophysics, 60 Garden Street, Cambridge, MA 02138, USA\\
  $^{11}$ Department of Physics and Astronomy, Stony Brook University, Stony Brook, NY, 11794, USA\\
  $^{12}$ Departament de F\'isica Aplicada, Universitat d'Alacant, Ap. Correus 99, 03080, Alacant, Spain\\
  $^{13}$ INAF -- Osservatorio Astronomico di Brera, Via E. Bianchi 46, I-23807, Merate, Italy\\
  $^{14}$ AIM (CEA/DSM-CNRS-Universit\'e Paris Diderot), Irfu/Service d'Astrophysique, Saclay, F-91191 Gif-sur-Yvette, France\\
  $^{15}$ ESAC, European Space Astronomy Centre, Villanueva de la Ca\~nada, E-28692 Madrid, Spain\\}

\def\src {SWIFT\,J1822.3-1606}
\def\xmm {\emph{XMM-Newton}}
\def\cxo {\emph{Chandra}}
\def\swift {\emph{Swift}}

\def\rxte {\emph{RXTE}}

\def\xte {XTE\,J1810--197}

\def\wes {CXOU J164710.2--455216}

\def\sgrlowb{SGR\,0418+5729}
\def\sgrswift{SWIFT\,J1822.3-1606}

\newcommand{\suzaku}{{\em Suzaku}}
\newcommand{\fermi}{{\em Fermi}}
\newcommand{\integral}{{\em INTEGRAL}}
\newcommand{\bc}{\begin{center}}
\newcommand{\ec}{\end{center}}

\def\lum {\mbox{erg s$^{-1}$}}
\def\nh {$N_{\rm H}$}
\def\ss {\mbox{s~s$^{-1}$}}
\def\cm2 {\mbox{cm$^{-2}$}}

\def\ltsima{$\; \buildrel < \over \sim \;$}
\def\lsim{\lower.5ex\hbox{\ltsima}}
\def\loe{\lower.5ex\hbox{\ltsima}}
\def\gtsima{$\; \buildrel > \over \sim \;$}
\def\gsim{\lower.5ex\hbox{\gtsima}}
\def\goe{\lower.5ex\hbox{\gtsima}}
\def\ergscm2 {erg\,s$^{-1}$cm$^{-2}$}
\def\ss {s\,s$^{-1}$}
\def\cm2 {cm$^{-2}$}

\def\ergs {${\rm erg\, s}^{-1}$}

\begin{document}
\date{MNRAS, accepted October 2015}
\maketitle

\begin{abstract}

We study the timing and spectral properties of the low-magnetic field, 
transient magnetar \src\ as it approached quiescence. 
We coherently phase-connect the observations over a time-span of $\sim$500 days since the discovery 
of \sgrswift\ following the \swift-BAT trigger on 2011 July 14, 
and carried out a detailed pulse phase spectroscopy along the outburst decay. 
We follow the spectral evolution of different pulse phase intervals and
find a phase and energy-variable spectral feature, which we interpret as proton cyclotron 
resonant scattering of soft photon from currents circulating in a strong 
($\gtrsim 10^{14}$ G) small-scale component of the magnetic field near the neutron star surface,
superimposed to the much weaker ($\sim 3 \times 10^{13}$ G)  magnetic field. We discuss also
the implications of the pulse-resolved spectral analysis for the emission regions on the surface 
of the cooling magnetar.

\end{abstract}


\section{Introduction}

Magnetars are isolated neutron stars believed to possess very strong magnetic fields 
($\sim$ $10^{14}$ -- $10^{15}$\,G; \citealt{duncan92,thompson95}) which power  
their bright X-ray emission, as well as their occasional outbursts and flares.
 The magnetar class is historically composed of two classes of sources, the Anomalous 
X-ray Pulsars (AXPs) and Soft $\gamma$-ray Repeaters (SGRs), which share 
a wide range of characteristics, mainly (see \citealt{mereghetti08} for a review): 
\begin{itemize}
\item spin periods in the 2 -- 12\,s range;
\item large positive period derivatives ($10^{-13}$ -- $10^{-10}$ \ss);
\item X-ray luminosities in the range $10^{33}$ -- $10^{35}$ \lum;  
\item sporadic bursting activity on timescales from ms to minutes.
\end{itemize} 

While a decade ago their persistent X-ray emission (outside their bursting periods) 
was believed to be steady. In 2004 \citet{ibrahim04} discovered the first transient 
magnetar, XTE J1810-197. The source was observed at a peak persistent luminosity 
$\sim 100$ higher that the quiescent value ($\sim$$10^{33}$ \lum; \citealt{gotthelf04}). 
Other magnetars have since been discovered which undergo transient outbursts lasting 
months to years until the quiescence is recovered. Therefore many unidentified faint 
X-ray sources might host quiescent magnetars that can be identified as they become 
active. Thanks to the observing capabilities of present space missions such as \swift, 
\integral\ and \fermi, several new magnetars have been discovered through the detection 
of short bursts events (10-100 ms duration) and/or long-term (months--years) outbursts 
(see \citealt{rea11} for a review). 

The magnetic field strength of magnetars is usually estimated by  assuming that, like 
ordinary pulsars, they spin down mainly through magnetic dipole losses. The intensity 
of the dipole magnetic field at the star equator is estimated as:  
$B_{dip} \sim 3.2 \times 10^{19} (P \dot{P})^{1/2}$ G, 
where $P$ is the spin period in seconds, $\dot{P}$ its first time derivative and a 
neutron star with radius $R \sim 10^6$cm and moment of inertia $10^{45}$ g cm$^2$ is
assumed.

Until some years ago all such measurements led to dipolar field strengths of 
 $\sim 10^{14}$--$10^{15}$\,G. However, the monitoring of the outburst 
of \sgrlowb\ demonstrated the existence of sources showing typical magnetar-like 
outbursts and short bursts, but with dipole fields
in line with ordinary pulsars (i.e. $\sim6\times10^{12}$\,G; \citealt{Rea13}). 
Simulations of magnetic field evolution in neutron stars 
with different initial field strength and configuration have shown that, a 
relatively old magnetar such as \sgrlowb\, despite its low surface dipolar 
field, might still harbor a sufficiently intense internal/crustal toroidal field 
to give rise to outbursts and short X-ray bursts (\citealt{Turolla11}).

Two more low-field magnetars were recently discovered thanks to their outburst activity, 
3XMM J185246.6$+$003317 (\citealt{Zhou14}, \citealt{Rea14}) and \src\, 
(Rea et al. 2012; Scholtz et al. 2012). The latter source was studied through 
the long-term monitoring that we present in this work.

\subsection{SWIFT J1822.3-1606}
The magnetar \sgrswift\ (Swift J1822, hereafter) was discovered 
through the detection of a series of bursts by the \swift\ Burst 
Alert Telescope (BAT) and \fermi\ Gamma-ray Burst Monitor (GBM) 
(\citealt{cummings11}) in July 2011. Soon afterwards \citet{pagani11} 
found the position of the new source at RA (J2000): 18h 22m 18.00s 
Dec (J2000): -16d 04$'$ 26.8$''$, with 1.8 arcsec radius of uncertainty 
(90\% confidence). Subsequently, the source was followed by virtually 
all the current generation of X-ray satellites. The results of the first 
9 months of X-ray monitoring of this new magnetar were presented and 
discussed by \citeauthor{rea12} (2012; hereafter R12), \citet{scholz12} 
and \citet{Scholz14}. In this work we perform an unprecedent coherent, 
pulse phase resolved spectroscopy (PPS) analysis spanning more than 400 
days of outburst decay (see Fig. \ref{phase}).

\section{Observations and Data Processing}
For the study of the outburst decay we used data from one 
\xmm, four \cxo\ and ten \emph{Swift} observations in 
addition to the data used in R12. We also used the \cxo\ 
ACIS-S pointings carried out at the beginning of the outburst 
in order to perform a detailed time-resolved PPS study. 
A log of the data collected during 2012 is given in Table 
\ref{obs-log}.\\

\begin{table}
\centering
\caption{Observations used in the timing analysis.
In bold are denoted the observations added to those used in R12.}
\label{obs-log}
{\footnotesize
\begin{tabular}{@{}cccc}
\hline
\hline
Instrument & Obs.ID & Date$^{a}$ & Exposure\\
 & & (MJD TBD) & (ks)\\
\hline
\swift & 00032033001 (PC) & 55757.75058 & 1.6 \\
\rxte  & 96048-02-01-00   & 55758.48165& 6.5 \\
\swift & 00032033002 (WT) & 55758.68430 & 2.0 \\
\swift & 00032033003 (WT) & 55759.69082 & 2.0 \\
\rxte  & 96048-02-01-05   & 55760.80853 & 1.7 \\
\swift & 00032033005 (WT) & 55761.54065 & 0.5 \\
\rxte  & 96048-02-01-01   & 55761.55969 & 5.0 \\
\swift & 00032033006 (WT) & 55762.24089 & 1.8 \\
\rxte  & 96048-02-01-02   & 55762.47384 & 4.9 \\
\swift & 00032033007 (WT) & 55763.30400 & 1.6 \\
\rxte  & 96048-02-02-00   & 55764.61846 & 6.1 \\
\rxte  & 96048-02-02-01   & 55765.46687 & 6.8 \\
\swift & 00032033008 (WT) & 55765.85252 & 2.2 \\
\swift & 00032033009 (WT) & 55766.28340 & 1.7 \\
\rxte  & 96048-02-02-02   & 55767.59064 & 3.0 \\
\rxte  & 96048-02-02-03   & 55769.35052 & 3.4 \\
\swift & 00032033010 (WT) & 55769.49531 & 2.1 \\
\swift & 00032033011 (WT) & 55770.39936 & 2.1 \\
\cxo       & 13511 & 55770.83049 & 11.7 \\
\swift & 00032033012 (WT) & 55771.23302 & 2.1 \\
\rxte  & 96048-02-03-00   & 55771.34185 & 6.8 \\
\swift & 00032033013 (WT) & 55772.40044 & 2.1 \\
\rxte  & 96048-02-03-01   & 55774.34999 & 6.9 \\
\bf \cxo & 12613& 55777.22193& 15.0\\
\rxte  & 96048-02-03-02   &  55777.85040 & 1.9 \\
\swift & 00032051001 (WT) & 55778.10744 & 1.7 \\
\swift & 00032051002 (WT) & 55779.18571 & 1.7 \\
\rxte  & 96048-02-04-00   &  55780.85040 & 6.7 \\
\swift & 00032051003 (WT) & 55780.49505 & 2.3 \\
\swift & 00032051004 (WT) & 55781.49878 & 2.3 \\
\rxte  & 96048-02-04-01   &  55782.57749 &  6.2\\
\rxte  & 96048-02-04-02   &  55784.97179 &  6.2\\
\swift & 00032051005 (WT) & 55786.42055 & 2.2 \\
\swift & 00032051006 (WT) & 55787.58688  & 2.2 \\
\rxte  & 96048-02-05-00   &  55788.05419 &  6.0\\
\swift & 00032051007 (WT) & 55788.25617  & 2.3 \\
\swift & 00032051008 (WT) & 55789.66173  & 1.7 \\
\rxte  & 96048-02-05-01   &  55789.95880 &  6.0\\
\swift & 00032051009 (WT) & 55790.36270  & 2.2 \\
\rxte  & 96048-02-06-00   &  55794.45899 &  6.5\\
\rxte  & 96048-02-07-00   &  55799.61550 &  6.9\\
\swift & 00032033015 (WT) & 55800.86278  & 2.9\\
\swift & 00032033016 (WT) & 55807.48660  & 2.4\\
\rxte  & 96048-02-08-00   &  55810.37979 &  6.0\\
\suzaku/XIS & 906002010 & 55817.92550 & 33.5 \\
\rxte  & 96048-02-10-00   &  55820.23970 &  6.7\\
\bf \cxo & 12614& 55822.79364& 10.1\\
\swift & 00032033017 (WT) & 55822.82836  & 4.9\\
\swift & 00032033018 (WT) & 55824.71484  & 1.5\\
\rxte  & 96048-02-10-01   & 55826.18540 & 5.6 \\
\xmm & 0672281801 & 55827.25350 & 10.6 \\
\swift & 00032033019 (WT) & 55829.45421  & 2.3\\
\swift & 00032033020 (WT) & 55835.54036  & 2.6\\
\rxte  & 96048-02-11-00   & 55835.90370 & 7.0 \\
\swift & 00032033021 (WT) & 55842.06040  & 4.2\\
\rxte  & 96048-02-12-00   & 55842.23269 & 5.8\\
\xmm & 0672282701 &  55847.06380 & 25.8 \\
\swift & 00032033022 (WT) & 55849.61916  & 3.4\\
\rxte  & 96048-02-13-00   & 55849.6597976 & 5.6 \\ 
\swift & 00032033024 (WT) & 55862.59155  & 10.2\\
\rxte  & 96048-02-14-00   & 55863.11100 & 5.6 \\ 
\hline
\end{tabular}}
\begin{list}{}{}
\item[$^{}$] Continues below.
\end{list}
\end{table}
\setcounter{table}{0}
\begin{table}
\centering
\caption{Continuation}
{\footnotesize
\begin{tabular}{@{}cccc}
\hline
\hline
Instrument & Obs.ID & Date$^{a}$ & Exposure\\
 & & (MJD TBD) & (ks)\\
\hline
\bf \cxo & 12615& 55867.17461& 16.3 \\ 
\swift & 00032033025 (PC) & 55977.16600& 6.3 \\
\swift & 00032033026 (WT) & 55978.53399 & 10.2 \\
\swift & 00032033027 (PC) & 55981.99499  & 11.0 \\
\swift & 00032033028 (WT) & 55982.96299   & 7.0\\
\swift & 00032033029 (WT) & 55985.17799 & 7.0 \\
\swift & 00032033030 (WT) & 55985.55000  & 7.0 \\
\swift & 00032033031 (WT) & 55991.09231 & 6.7 \\
\xmm & 0672282901&  56022.95692 & 26.9 \\
\swift & 00032033032 (WT) & 56031.141159 & 4.3 \\
\bf \cxo & 14330 & 56037.08846 & 20.1 \\
\bf \swift& 00032033033 & 56052.66948& 5.2\\
\bf \swift& 00032033034 & 56073.25354& 4.9\\
\bf \swift& 00032033035 & 56095.56083& 5.6\\
\bf \swift& 00032033036 & 56104.55201& 6.2\\
\bf \swift& 00032033037 & 56114.31112& 6.8\\
\bf \swift& 00032033038 & 56136.32359& 9.0\\
\bf \swift& 00032033039 & 56156.20958& 4.9\\
\bf \swift& 00032033040 & 56161.69589& 5.0\\
\bf \xmm  & 0672283001& 56178.85111  & 21.7 \\
\bf \swift& 00032033042 & 56206.00752& 5.0\\
\bf \swift & 00032033043  & 56238.70919 & 4.9 \\
\hline
\end{tabular}}
\begin{list}{}{}
\item[$^{a}$] Mid-point of the observations.
\end{list}
\end{table}



The data reduction was performed by means of the standard 
procedures outlined in R12: it consists of initial raw data 
calibration, filtering from soft-proton flares, correcting 
the arrival times to the barycenter of the solar system, 
source and background extraction, pileup checks, spectral 
data rebinning and grouping (see Sec. \ref{ppsana} for datails).
These steps were performed by using the official Science
Analysis System (\textsc{sas}) package (version 11) for 
the \xmm\ data, and the Chandra Interactive Analysis of Observations 
(\textsc{ciao}) system (version 4.4) for the \cxo\ data. 
The \swift\ data were processed and filtered with standard 
procedures and quality 
cuts\footnote{See http://swift.gsfc.nasa.gov/docs/swift/analysis/ for more details.} 
using \textsc{ftools} tasks in the \textsc{heasoft} 
software package (v. 6.12) and the calibration files from the latest 
\textsc{caldb} release.  

For the spectral analysis we used \textsc{xspec} (version 12.7.1), 
for the timing analysis, \textsc{xronos} (version 5.22) and pipelines developed 
in-house for the phase fitting procedures.

\begin{figure}
  \centering
  \includegraphics[width=0.35\textwidth, angle=270]{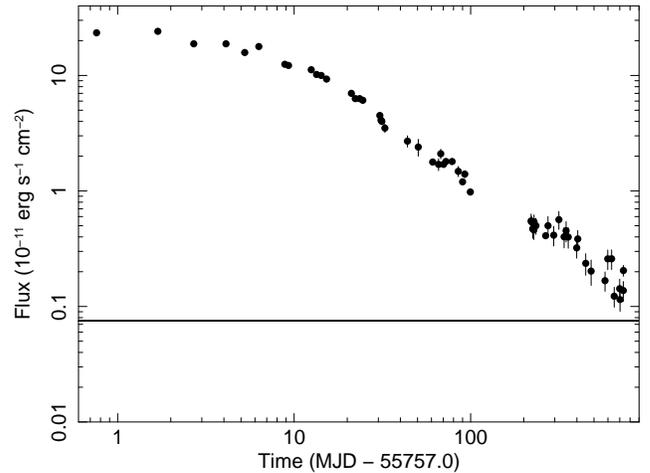}
  \caption[Flux decay of Swift J1822.3--1606 in the 1--10 keV energy
    range]{Swift-RXT flux decay of Swift J1822.3--1606 in the 1--10 keV energy range.
    The solid line represents the quiescence flux level.}
  \label{fluxdec}
\end{figure}

\begin{figure}
    \centering
    \includegraphics[width=.42\textwidth, angle=270]{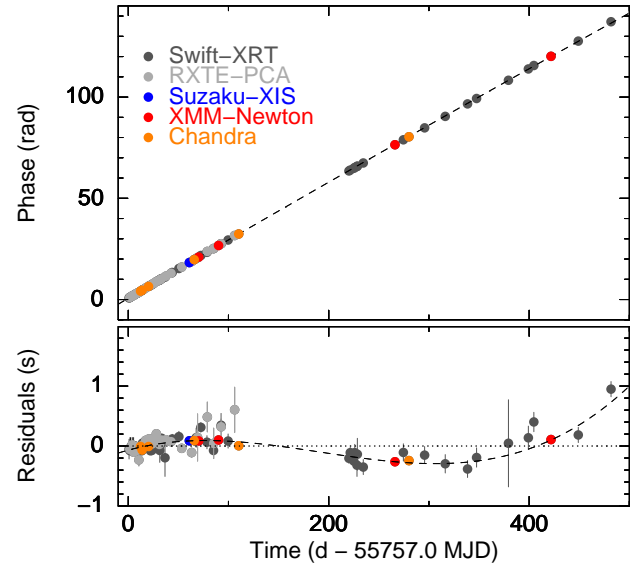}
    \caption[Pulse phase fitting]
            {Pulse phase evolution over time.
              In the lower panel are shown the time residuals after correcting the 
              linear and quadratic components (correction to the $P$ and $\dot{P}$ values). 
              The dashed line in the residual panel marks the detected {\it \"{P}} component.
              The epoch of reference is 55757.0 Modified Julian Date (MJD). See the text for
              details on the ephemerides.
            }
            \label{phase}
\end{figure}

\subsection{Timing}
We begun by extending the validity of the phase-coherent timing 
solution reported by  R12 by adding  the new  datasets  listed in
Table\,\ref{obs-log}. In fact, the accuracy and time span between
the latest R12 observation and the first of the additional ones listed in
Table\,\ref{obs-log} ($\sim21$ days) is such that we do not miss any
cycle in the phase-fitting procedure. 

In order to fit the additional data, we added a second period derivative 
to the timing solution (see Fig. \ref{phase}, lower panel).
The introduction of the higher order period derivative results in a
significant improvement of the fit; an F-test gave a probability of
$>10\sigma$ that the inclusion of a cubic component is required for
our data set (see Fig. \ref{phase}).

\begin{table}
\caption{Timing parameters for \src.}
\begin{center}
{\small
\begin{tabular}{lr}
\hline
\hline
Reference Epoch (MJD) & 55757.0\\
Validity period (MJD) &  55757-- 56239\\
$P$\,(s) & $8.437720019(7)$ \\
$\dot{P}$\,(s\,s$^{-1}$) & $1.34(1)\times 10^{-13}$ \\
$\ddot{P}$\,(s\,s$^{-2}$) & $-5.1(2)\times10^{-21}$ \\
$\nu$\,(Hz) & $0.1185154281(1)$ \\
$\dot{\nu}$(s$^{-2}$)  & $-1.88(2)\times10^{-15}$  \\
$\ddot{\nu}$(Hz\,s$^{-2}$)  & $7.1(2)\times10^{-23}$\\
$\chi^2/$dof & 168/81 \\
RMS residuals (ms) & 360 \\
\hline
B (Gauss) & $\sim 3.4\times10^{13}$ \\
L$_{\rm rot}$ (erg~s$^{-1}$) & $\sim 8.9\times
10^{30}$ \\
$\tau_{\rm c}$ (Myr) &  $\sim 1.0$ \\
\hline
\hline
\end{tabular}}
\end{center}
\label{tab:timing}
\end{table}

The best timing solution for our data set is: $P=8.437720019(7)$ s,
$\dot{P}=1.34(1)\times10^{-13}$s\,s$^{-1}$ and {\it \"P}
$=-5.1(2)\times10^{-21}$s\,s$^{-2}$ (1$\sigma$ c.l., 3 parameters of
interest for epoch 55757.0 MJD; see also Table\,\ref{tab:timing}). 
Based on the best fitting $P$ and $\dot{P}$, the inferred surface dipolar magnetic field is 
$B \sim 3.4(1)\times10^{13}$ G at the equator. 
This value lies in between previous results of \citet{scholz12}: $B \sim 5\times10^{13}$ G, 
and  R12: $B \sim 2.7\times10^{13}$ G, and is higher than the estimate
of \citet{Scholz14}: $B \sim 1.4\times10^{13}$ G. Yet, it is lower
then the B$_{QED}$ critical value ($B \sim 4.4\times10^{13}$ G): \src\ is thus 
the magnetar with the second lowest magnetic field. 
This timing solution also implies a 
spin-down power L$_{\rm rot}=4\pi^2 I \dot{P}/P^3\sim 9\times 10^{30}$ \ergs,  
assuming uniform-density neutron star with moment of inertia $I = 10^{45}$ g cm$^2$.
The inclusion of a third period derivative, or other components, 
such as a post glitch recovery, does not significantly improve 
the fit. However, the timing solution by \citet{Scholz14} over a 
longer -- with respect to this work -- time span of $\sim$ 2.3 years 
did indeed improve with the inclusion 
of an exponential glitch recovery for the first $\sim$ 60 days, and a 
separate $P$-$\dot{P}$ timing solution for the rest of the data set, 
see \citet{Scholz14} for details on their solution. For the purpose of this work a single 
coherent timing solution for the entire data set is desirable and we 
use our timing solution to perform the pulse phase-resolved analysis.  
We note that the timing solutions with parameters 
closer to ours are those reported by R12 and the 2nd solution of 
\citet{scholz12}, which were obtained over a time-span of $\sim$ 1 year.

\begin{figure}
  \centering
  \includegraphics[width=0.9\textwidth, angle=-90]{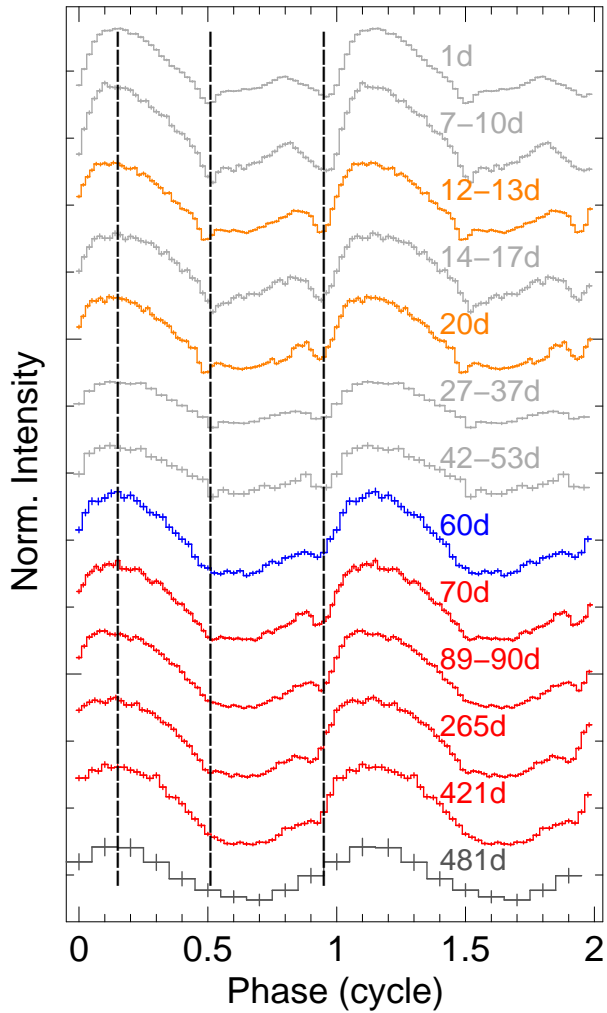}
  \caption{Selected pulse shapes plotted as a function of time for
    different epochs of the 2011-2012 outburst of \src. The 1-10keV
    light curves (2-10keV for RXTE) have been folded to the $P -
    \dot{P} - ${\it \"P} phase-coherent solution reported in
    Table\,\ref{tab:timing}. Different colors mark data from different
    missions similarly to the color code used in
    Figure\,\ref{phase}. }
  \label{timing}
\end{figure}

\subsection{Spectral Analysis}
\label{specana}
\subsubsection{Phase-average analysis}
\label{argv}


Phase-averaged spectral analysis were performed by R12,  
\citet{scholz12} and \citet{Scholz14}. In all cases a model composed by a 
photoelectrically absorbed blackbody (BB) plus a power-law (PL) was used
to fit the spectra; in R12 a two-BB was used as well. 
In both \citet{scholz12} and  \citet{Scholz14}, acceptable fits 
were obtained in the $1 - 10$ keV band. In R12 over the $0.6 - 10$ keV 
energy band (for both models).
Restricting the \xmm\ data to those energy intervals, we were 
able to reproduce their results for all the common \xmm\ spectra. 
However, the situation changes when we considered a larger 
energy range, $0.3 - 10$ keV.

\begin{figure}
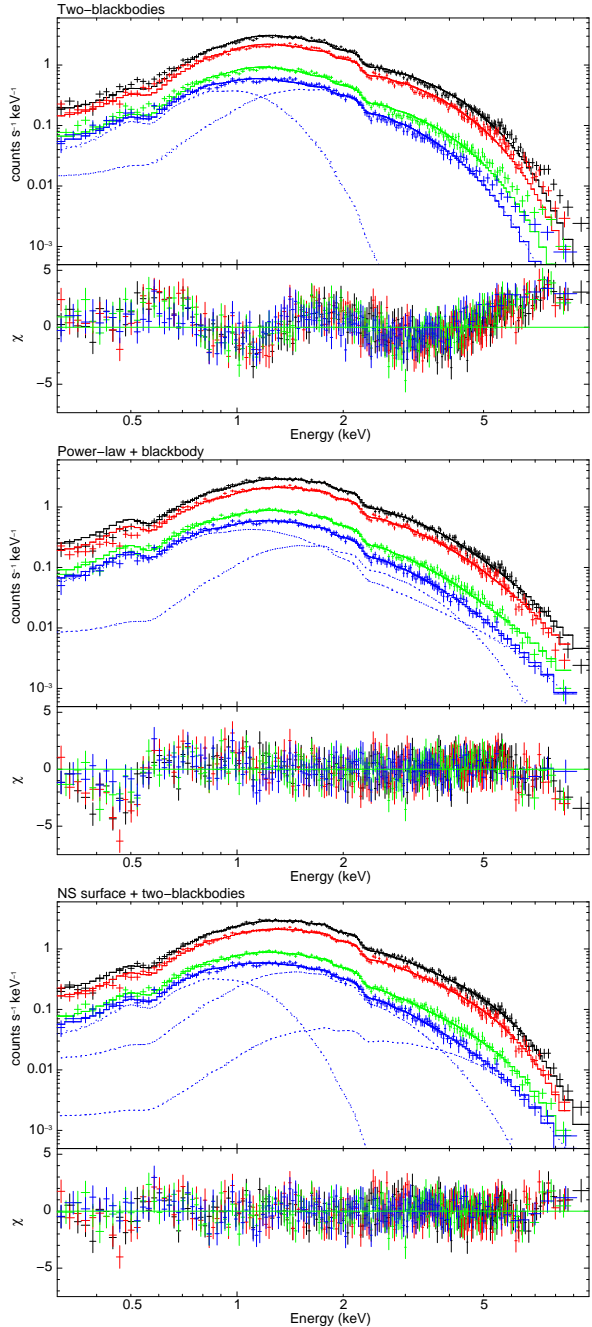

  \centering
  \includegraphics[width=0.33\textwidth, angle=270]{2bbcomps.eps}
  \includegraphics[width=0.33\textwidth, angle=270]{plcomps.eps}
  \includegraphics[width=0.33\textwidth, angle=270]{3bbcomps.eps}
  \caption{\src\ \xmm\ spectra. Best fits using 2-blackbodies (upper panel),
    blackbody plus power-law (middle panel), and 2+1-blackbodies
    (lower panel). $\chi^2_{red} = 2.31, 1.44$ and $1.08$, respectively.
    The fits components of the last observations (blue dotted lines)
    are shown for comparison. Note that by the last observation the 
    source has significantly softened and the hot BB component is not 
    dominant. See the text for details.}
  \label{otherfits}
\end{figure}

Indeed, the two-BB model failed to fit the high energy part 
($> 6$ keV) of the spectra (see e.g. Fig. \ref{otherfits}, 
upper panel, where a $\chi^2_{red} = 2.31$ for 674 
degrees of freedom, dof was obtained).
Following the same rationale as in the case of \wes\ 
(\citealt{rodriguez13}), we added an additional (``cold'') 
blackbody with fixed temperature to the two-BB model. 
Thus, adding one additional parameter to the fit, namely, 
the cold blackbody radius. This additional component is 
meant to account for a colder surface region, which might 
correspond to either the rest of the NS surface, or a portion
of it.
The kT$_{BB} \sim$ 150 eV value is based on spectral analysis of the 
quiescent emission of \src, previously performed in R12 and \citet{scholz12}. 
Note that estimates for other magnetars (i.e. \xte\ and \wes, see 
\citealt{Albano10}) yield similar values.

The addition of the colder BB yields an improved $\chi^2_{red}$ of 1.08 (670 dof)
(see Fig. \ref{otherfits} and Table \ref{phavrg}).
We studied the significance of such component by calibrating the 
F-statistics using simulations of the null model (the double-blackbody
model) as suggested by \citet{Protassov02}. In accordance with this approach, 
the distribution of the null F-statistic was produced by fitting each simulated 
spectrum with both the null and the triple-blackbody model and extracting the 
relative F-statistic. Running $5\times10^5$ Monte Carlo simulations, we 
obtained that the chance occurence probability of such an improvement is lower
than $2\times10^{-6}$.

\begin{table*}
\begin{minipage}{110mm}
\renewcommand{\arraystretch}{1.3}
\resizebox{8cm}{!} {
\begin{minipage}{7.5cm}
\begin{tabular}{|l|l|c|c|c|c|}
\hline
\multicolumn{6}{|c|}{\src\  1+2-blackbody phase-average spectral fit} \\
\hline
Obs. MJD & T$_h$ (keV)  &    R$_{BB}^h$ (Km)  &    T$_w$ (keV)  &        R$_{BB}^w$ (Km) &   R$_{BB}^c$ (Km)    \\ \hline
15777  &  $1.07^{+0.04}_{-0.03}$   &  0.29   $\pm$ 0.02    &   0.52   $\pm$ 0.01  &  1.26 $\pm$ 0.04 &  11.6  $\pm$ 0.2  \\
15827  &  $0.87^{+0.04}_{-0.03}$   &  0.23   $\pm$ 0.02    &   0.41   $\pm$ 0.02  &  $0.96^{+0.05}_{-0.04}$ &  7.0  $\pm$ 0.3  \\
15846  &   0.83 $\pm$ 0.03       &  0.22   $\pm$ 0.01    &   0.41   $\pm$ 0.02 &  0.82   $\pm$ 0.04 &  6.3  $\pm$ 0.3  \\
16023  &   0.87 $\pm$ 0.03       &  $0.10^{+0.02}_{-0.01}$ &   0.42   $\pm$ 0.02  &  $0.54^{+0.02}_{-0.03}$ &  4.52 $\pm$ 0.2  \\
16178  &  $0.98^{+0.10}_{-0.08}$   & $0.05^{+0.02}_{-0.01}$  &   0.48    $\pm$ 0.02  &  $0.36^{+0.02}_{-0.03}$ &  4.2  $\pm$ 0.2  \\ \hline
\hline
\end{tabular}
\end{minipage}
}
\caption{Parameters of 1+2BB fit of \cxo\ (MJD 15777) and \xmm\ 
  spectra (See text for details on the model). 
  The distance used to calculate the R$_{BB}$ was 1.6 kpc, 
  as proposed by \citet{scholz12}. Error are 1-$\sigma$.}
\label{phavrg}
\end{minipage}
\end{table*}

Beside significantly improving the fits, the 1(T-fixed)+2(free) 
BB (1+2BB, for short) model provides consistent results.
First, the radius, which is the only free parameter of the 
additional, temperature-fixed BB, varies from $4.0\,d_{1.6}$ km to 
$7.3\,d_{1.6}$ km, where $d_{1.6}$ is the distance to the magnetar
in units of 1.6 kpc, the value proposed by \citet{scholz12} (see below). 
Another hint at the validity of the model is the fact that 
the measured radius of the cold component is consistent with 
the BB radius observed in quiescence, before the outburst,
at the same temperature (R12, \citealt{scholz12}). Note as well 
that the measured radius, both in our analysis as well as in 
the cited papers, is significantly less then the expected 
$\sim 10$ km for a neutron star, suggesting that it may not 
correspond to the whole surface.




On the other hand, the PL+BB model, while fitting well the data in 
the 0.6--10 keV range (R12), shows significant residuals at 
lower energies $< 0.6$ keV (see Fig. \ref{otherfits}, middle 
panel). In the 0.3--10 keV range the PL+BB model yields a
poor $\chi^2_{red}\,$/dof $ = 1.44\,$/674. In the case
of the PL+BB model, adding the same temperature-fixed BB, 
as it was done for the 2BB model, only marginally improves
the fit ($\chi^2_{red}\,$/dof $ = 1.22\,$/670).

Using the best 1(T-fixed)+2(free) BB model and the whole set 
of phase-average spectra, we obtained a 
best-fit column density value of $N_{\mathrm{H}} = 4.7(1) \times 10^{21}$ 
cm$^{-2}$, which is very close to the values found in previous works, 
R12 $N_{\mathrm{H}} = 5.0(1) \times 10^{21}$ cm$^{-2}$ 
; and \citet{scholz12} $N_{\mathrm{H}} = 4.5(1) \times 10^{21}$ cm$^{-2}$. 
Moreover, this value is consistent with the 
$N_{\mathrm{H}} = 4(1) \times 10^{21}$ cm$^{-2}$ 
of the Galactic H\textsc{ii} region 
M17 (\citealt{Townsley03}), located about 20$'$ from Swift J1822 
and at $1.6 \pm 0.3$ kpc from us (\citealt{Nielbock01}), as noted
by \citet{scholz12}. We kept $N_{\mathrm{H}}$ fixed at 
$4.7 \times 10^{21}$ cm$^{-2}$ 
for the rest of the analysis, and adopted a tentative distance of $1.6$ 
kpc in BB radii calculations.

For this analysis the solar abundances by \citet{Anders89} were used to account  
for the photoelectric absorption, the same used in all previous 
analysis (R12, \citealt{scholz12}). 
 In addition, we used the same photoelectric cross-sections by 
\citet{Balucinska-Church92} used in the cited works. 
We also tried other solar abundances (\citet{Feldman92}; 
\citet{Wilms00}; \citet{Lodders03} and \citet{Asplund09}) and 
in all cases the 1+2BB model showed a similar statistical 
predominance in comparision to the PL+BB and the
2BB models as described above.






\subsubsection{Coherent phase-resolved analysis}
\label{ppsana}

In this section we describe a pulse phase resolved 
analysis as a function of the decaying flux by using one \cxo\ 
plus the four \xmm\ pointings covering the outburst decay from 
day 12 up to day 421. To do so, we use our updated phase-coherent 
timing solution to fold and extract the spectra.

\begin{figure}
  \centering
  \includegraphics[width=0.33\textwidth, angle=270]{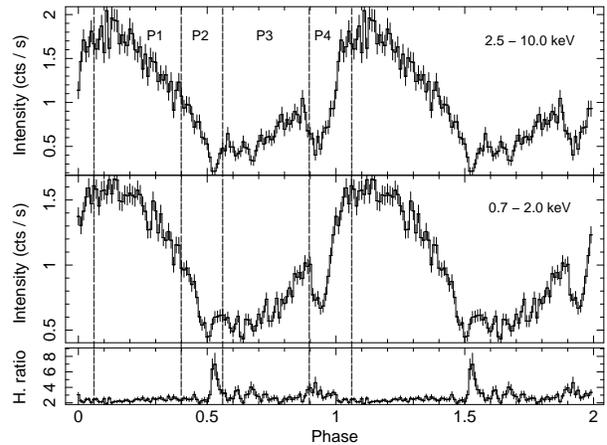}
  \caption{Pulse profile of \src\ in the 2.5--10 keV interval (upper panel);
    0.7--2 keV interval (middle panel) and hardness ratio for those intervals
    (lower panel). Obs id: 0672281801.
  }
  \label{hr}
\end{figure}

\begin{figure}
    \centering
      \includegraphics[width=.33\textwidth, angle=270]{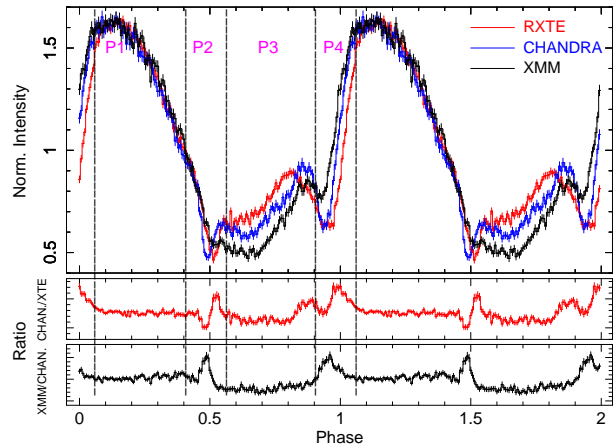}
    \caption{Upper panel: 0.3--10 keV Normalized pulse profiles 
      for the highest statistics light curves which are also a 
      function of increasing time and decreasing flux from the 
      outburst onset: \rxte, \cxo\ and \xmm\ in the order 
      of decreasing flux (see also Fig.\,\ref{timing}). 
      Lower panels: Ratios of the profiles used to identify the 
      phase interval with larger pulse profile variations.}
    \label{tratio}
\end{figure}

In order to define the phase interval selection we inspected spectral 
changes across the pulse profile in single observations (see e.g. 
Fig. \ref{hr}). A pulse profile comparison 
as a function of time is shoen in Fig. \ref{tratio}. In both figures similar
features are seen in the same phase regions, near phase 0.5 and 0.93.
Based on this and in order to make sure that each phase interval
contains enough counts, we adopted the phase intervals marked as P1, P2, P3 
and P4 in Fig. \ref{tratio}, which correspond to phases 
0.05-0.4, 0.4-0.55, 0.55-0.9 and 0.9-1.05, respectively.

For the spectral fitting (with \textsc{xspec} version 12.7.1), 
the data were grouped so as to have at least 30 counts per 
energy bin, while the instrument energy resolution were 
oversampled by a factor of three. The standard routines 
of \textsc{xmm-sas} and \textsc{ciao} were used to produce 
the ancillary response files and redistribution matrix files.

\begin{figure}
  \centering
  \includegraphics[width=0.5\textwidth, angle=0]{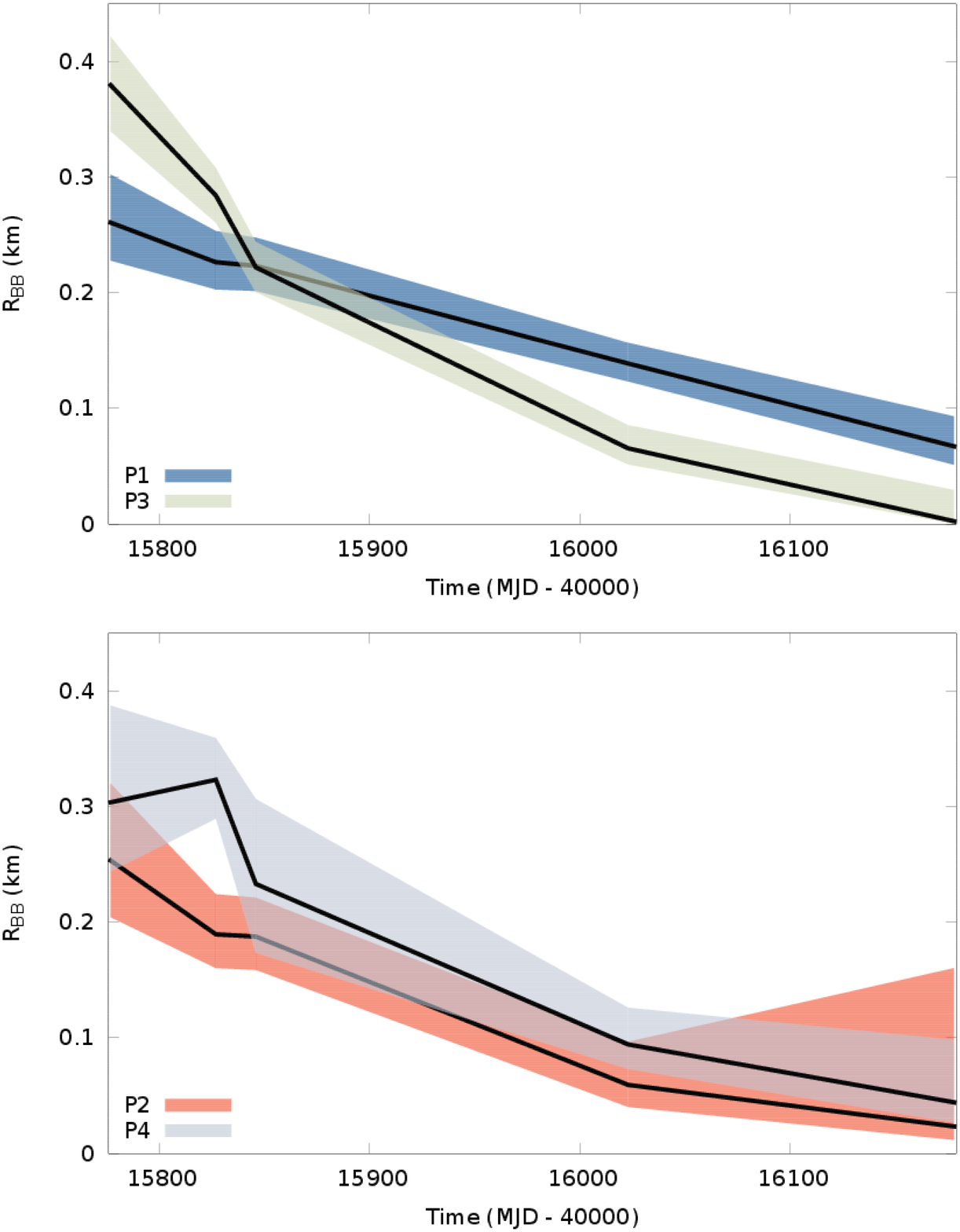}
  \caption{Evolution of the hot blackbodies radii ($R_{BB}$) used for modeling the pulse phase intervals, 
    see the text for details. The shadows represent the 1-$\sigma$ confidence interval.}
  \label{pps2.1}
\end{figure}

\begin{figure}
  \centering
  \includegraphics[width=0.5\textwidth, angle=0]{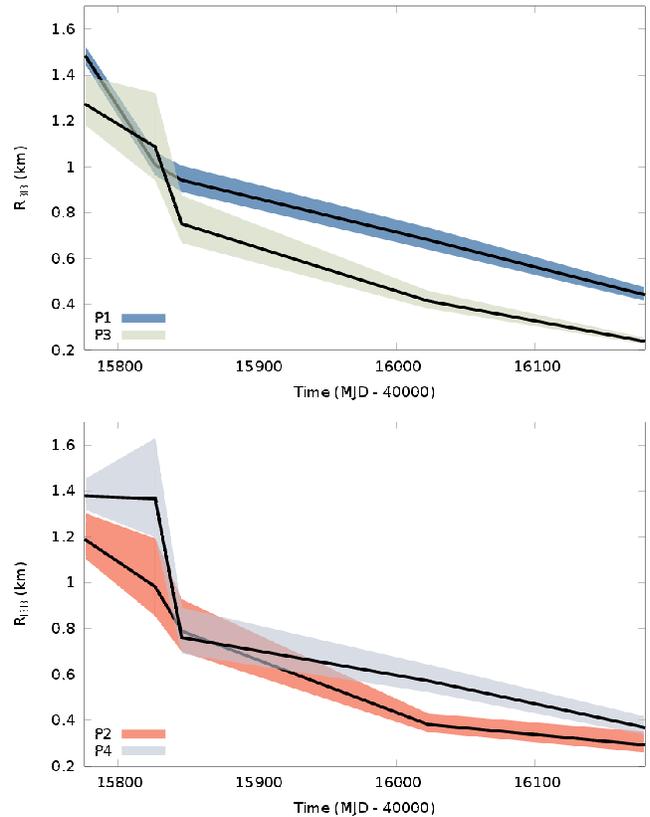}
  \caption{Same as Fig. \ref{pps2.1}, for the warm blackbodies.}
  \label{pps2.2}
\end{figure}

\begin{figure}
  \centering
  \includegraphics[width=0.5\textwidth, angle=0]{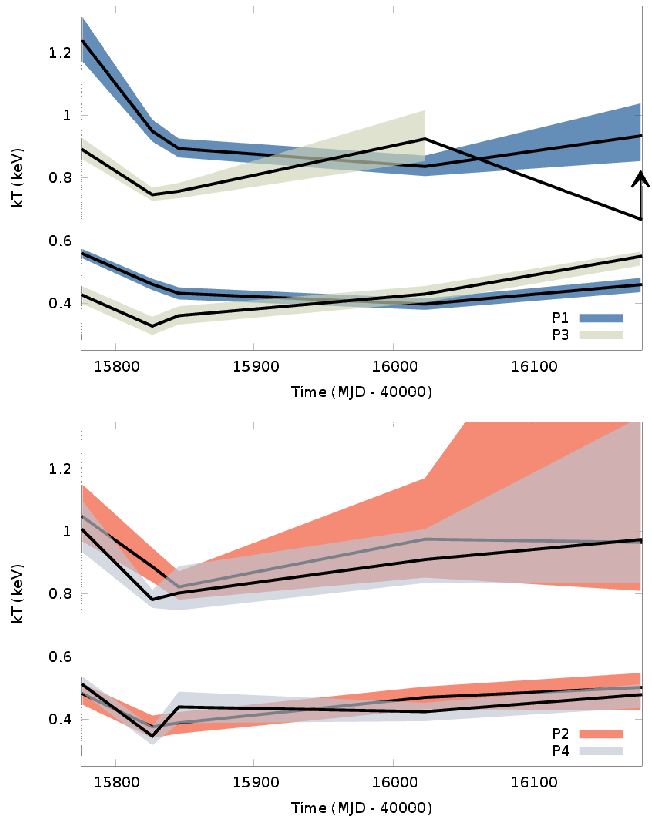}
  \caption{Temperature evolution of the blackbodies used for modeling the pulse phase intervals.
    Lower values correspond to the warm components, higher values to the hot components. 
    The shadows represent the 1-$\sigma$ confidence interval.
  }
  \label{pps2.3}
\end{figure}

For the phase-resolved spectroscopy, we used the 2(free)+1(T-fixed) BB 
model described above (Section \ref{argv}). We also assume that 
the geometric configuration of the emission zones does not vary 
dramatically during the outburst decay, so we follow each evolving 
component during the outburst decay. The observed pulse profile
relative stability over time (see e.g. Fig. \ref{tratio}) indicates 
that these assumptions are reasonable. In the context of our spectral
analysis, we interpret the flux evolution (Fig. \ref{fluxdec}) as 
changes in the radii and temperature of the two free blackbodies.
Since no geometric information about the specific configuration 
of the emission zones is available, the two blackbodies were 
left free to vary from phase to phase, in the case that more then 
one hot region is present. 
We favor a two-hotspot (emission zones) model (see Section \ref{discussion}), 
since a single hotspot configuration does not reproduce the observed pulse profiles.
The results are presented in Table \ref{1822pps} and graphically in Figs. \ref{pps2.1}, 
\ref{pps2.2} and \ref{pps2.3}, where phase intervals P1 and P3 are shown in the
upper panels and P2 and P4 in the lower panels. The phase intervals have been divided 
in this way for visualization purposes and because of similarities found in some intervals,  
in particular, between P2 and P4, which show quite similar values and evolution throughout
the outburst decay (see Section \ref{discussion}).


\begin{table*}
\begin{minipage}{150mm}
\renewcommand{\arraystretch}{1.3}
\resizebox{8cm}{!} {
\begin{minipage}{7.3cm}
\begin{tabular}{|l|l|c|c|c|c|c|}
\hline
\multicolumn{7}{|c|}{\src:  1+2-blackbody spectral fit} \\
\hline
Pulse phase interval\footnote{See Fig. \ref{tratio} for reference.}& Obs. MJD & T$_h$ (keV)  &    R$_{BB}^h$ (km)  &    T$_w$ (keV)  &   R$_{BB}^w$ (km)&   R$_{BB}^c$ (km)\footnote{Different solar abundances can reduce this value up to a factor of $\sim 2$. See alsoSection \ref{argv}} \\ \hline
\multirow{5}{*}{P1}      & 15777  &  $1.24^{+0.08}_{-0.07}$   & $0.26^{+0.04}_{-0.03}$   &  $0.56^{+0.01}_{-0.02}$    &  1.48   $\pm$ 0.04    &14.5$\pm$ 0.3\\
                         & 15827  &  $0.95^{+0.04}_{-0.03}$   & $0.23^{+0.03}_{-0.02}$   &   0.46   $\pm$ 0.02      &  1.01   $\pm$ 0.05    &8.9 $\pm$ 0.2 \\
                         & 15846  &   0.89 $\pm$ 0.03       &  0.22   $\pm$ 0.02      &   0.43   $\pm$ 0.02     &  $0.94^{+0.06}_{-0.05}$ &7.8 $\pm$ 0.2\\
$\chi^2_{red} = 1.1865$   & 16023  &   0.84 $\pm$ 0.03       &  0.14   $\pm$ 0.02      &   0.44   $\pm$ 0.02     &  $0.68^{+0.05}_{-0.04}$ &5.1 $\pm$ 0.1\\
                         & 16178  &  $0.94^{+0.10}_{-0.08}$   & $0.07^{+0.03}_{-0.02}$   &   0.46   $\pm$ 0.02      &  0.44   $\pm$ 0.03    &4.6 $\pm$ 0.1\\ \hline
\multirow{5}{*}{P2}      & 15777  &  $1.05^{+0.10}_{-0.08}$   & $0.25^{+0.05}_{-0.07}$   &   0.48   $\pm$ 0.03      &  $1.19^{+0.12}_{-0.08}$ &11.2$\pm$0.4\\
                         & 15827  &  $0.89^{+0.06}_{-0.05}$   &  0.19   $\pm$ 0.03     &   0.38   $\pm$ 0.04      &  $1.0^{+0.2}_{-0.1}$    &6.9$\pm$0.4\\ 
                         & 15846  &  $0.82^{+0.05}_{-0.04}$   &  0.19   $\pm$ 0.03     &   0.39   $\pm$ 0.03      &  $0.79^{+0.14}_{-0.09}$  &6.0$\pm$0.3\\ 
$\chi^2_{red} = 1.0522$   & 16023  &  $1.0^{+0.2}_{-0.1}$      & $0.06^{+0.04}_{-0.02}$   &   $0.47^{+0.03}_{-0.04}$  &  $0.38^{+0.05}_{-0.04}$  &4.5$\pm$0.2\\
                         & 16178  &  $1.0^{+0.9}_{-0.2}$     & $0.02^{+0.13}_{-0.01}$   &   $0.50^{+0.05}_{-0.07}$  &  $0.29^{+0.06}_{-0.03}$   &4.1$\pm$0.2\\ \hline
\multirow{5}{*}{P3}      & 15777  &  $0.89^{+0.04}_{-0.03}$   &  0.38   $\pm$ 0.04      &   0.43  $\pm$ 0.03      &  $1.27^{+0.12}_{-0.09}$  &11.2$\pm$0.3\\ 
                         & 15827  &   0.75  $\pm$ 0.02      &  0.28   $\pm$ 0.02      &   0.33  $\pm$ 0.03      &  $1.1^{+0.2}_{-0.1}$     &6.0$\pm$0.3\\
                         & 15846  &  $0.76^{+0.03}_{-0.02}$   &  0.22   $\pm$ 0.02      &   0.36  $\pm$ 0.03      &  $0.75^{+0.12}_{-0.08}$  &5.6$\pm$0.2\\
$\chi^2_{red} = 1.0516$   & 16023  &  $0.93^{+0.09}_{-0.07}$   & $0.07^{+0.02}_{-0.01}$   &    0.43  $\pm$ 0.03      &  $0.41^{+0.04}_{-0.03}$  &4.5$\pm$0.1\\ 
                         & 16178  &  $> 0.67$              &$0.002^{+0.029}_{-0.002}$     &   $0.55^{+0.01}_{-0.03}$  &  $0.24^{+0.02}_{-0.01}$ &4.4$\pm$0.1\\ \hline
\multirow{5}{*}{P4}      & 15777  &  $1.00^{+0.09}_{-0.07}$   & $0.30^{+0.08}_{-0.06}$    &   $0.51^{+0.02}_{-0.03}$  &  $1.38^{+0.08}_{-0.06}$  &11.4$\pm$0.5\\
                         & 15827  &   0.78   $\pm$ 0.03     & $0.32^{+0.04}_{-0.03}$    &   0.35  $\pm$ 0.03      &  $1.4^{+0.3}_{-0.2}$     &7.0$\pm$0.5\\  
                         & 15846  &  $0.80^{+0.09}_{-0.06}$   & $0.23^{+0.07}_{-0.06}$    &   0.44  $\pm$ 0.05      &  $0.76^{+0.13}_{-0.07}$  &7.6$\pm$0.3\\
$\chi^2_{red} = 1.0516$   & 16023  &  $0.91^{+0.10}_{-0.07}$   & $0.09^{+0.03}_{-0.02}$   &    0.42  $\pm$ 0.03      &  $0.57^{+0.07}_{-0.05}$  &4.8$\pm$0.2\\ 
                         & 16178  &  $1.0^{+0.3}_{-0.2}$      & $0.04^{+0.05}_{-0.02}$    &   $0.48^{+0.03}_{-0.04}$  &  $0.37^{+0.05}_{-0.03}$  &4.5$\pm$0.2\\ \hline
\hline
\end{tabular}
\end{minipage}
}
\caption{1+2BB fit spectral parameters of the phase resolved spectroscopy. The distance used to calculate the R$_{BB}$ was 1.6 kpc, 
  as proposed by \citet{scholz12}. Reported error are 1-$\sigma$.}
\label{1822pps}
\end{minipage}
\end{table*}

In order to maintain coherence with previous works and with our phase-averaged 
analysis, we used the \citet{Anders89} solar abundances 
and corrected the photoelectric absorption with \citet{Balucinska-Church92} 
photoelectric cross-sections in the pulse-phase resolved 
spectroscopy as well. We note that since different solar abundances 
affect only the lower energies, only the radius of the fixed, cold  BB, 
is affected (reduced it by a factor of $\lesssim 2$). The warm and hot BB, 
(see Table \ref{1822pps}) remain unaffected within the 1-$\sigma$ confidence 
interval, when the solar abundances are varied (see also Section \ref{argv}).



\subsubsection{Phase-variable spectral features}
\label{phspft}

\begin{figure}
  \centering
  \includegraphics[width=0.5\textwidth, angle=0]{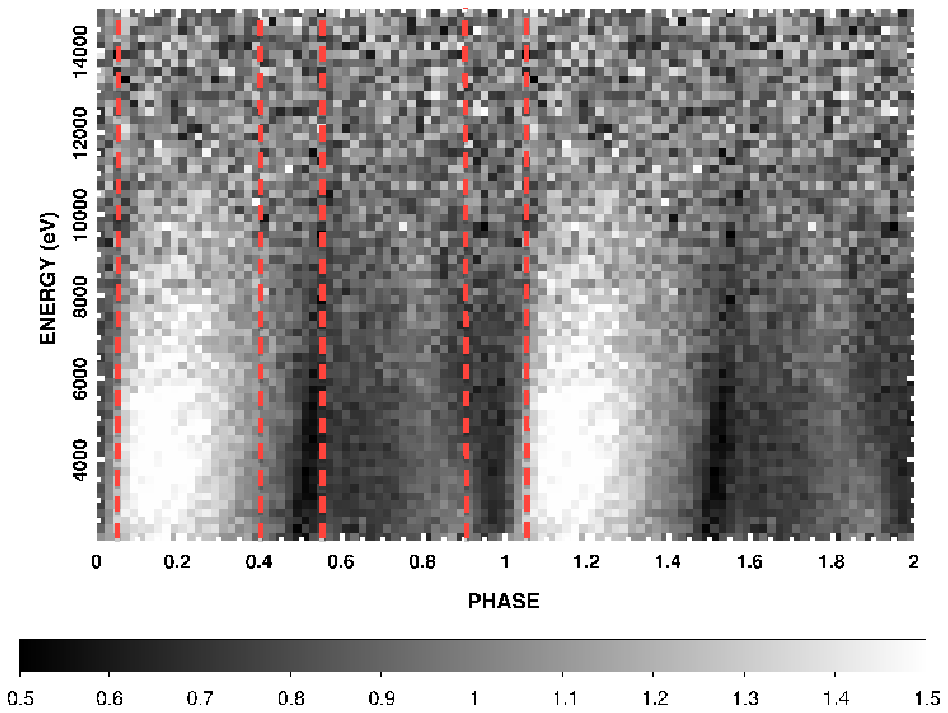}
  \includegraphics[width=0.5\textwidth, angle=0]{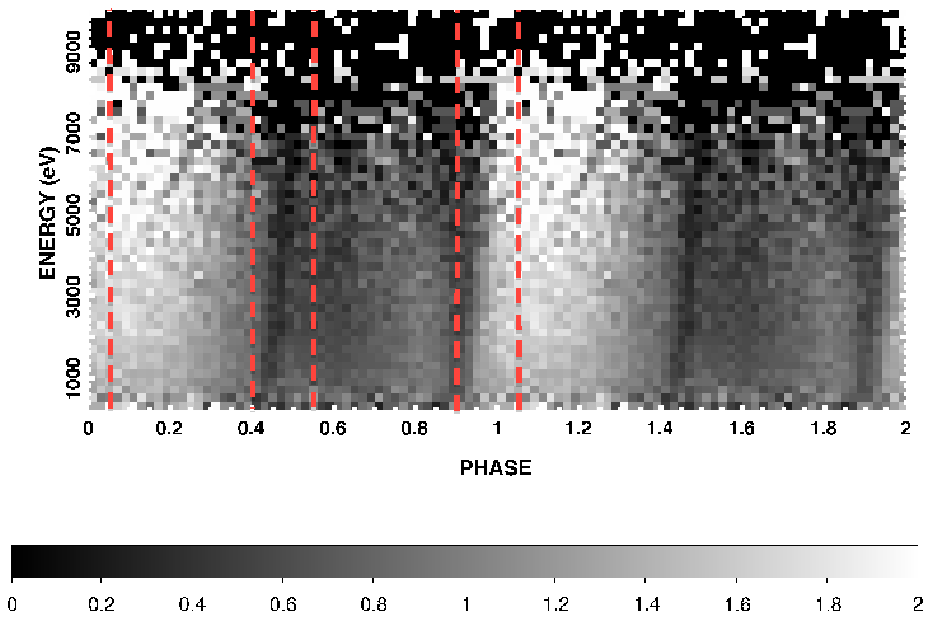}
  \includegraphics[width=0.5\textwidth, angle=0]{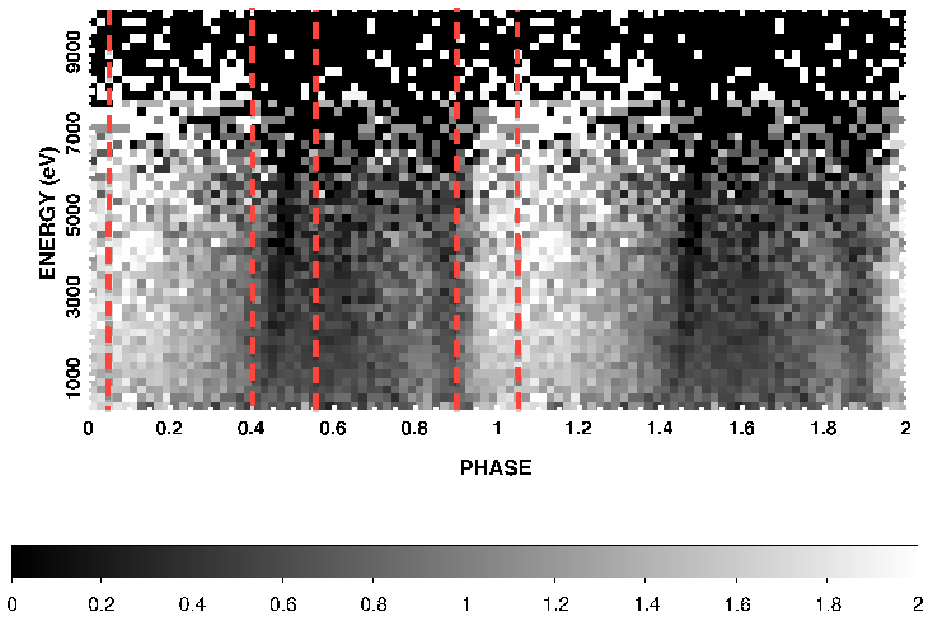}
  \caption{Phase-energy images, divided by the phase-average spectrum, of 
    \sgrswift\ from \rxte\  (PCA observations from 2011 July 16 to 2011 July 20, 
                    {\it top panel}), \cxo\ (ACIS-S observation in CC mode on 2011 July 27, 
                    {\it middle panel}) and \xmm\ (EPIC pn observation on 2011 September 23 
                    {\it bottom panel}) data. The vertical dashed lines denotate the intervals 
                    used in the PPS (see Fig. \ref{tratio}).
  }
  \label{PhaseEnergy}
\end{figure}

\begin{figure}
  \centering
  \includegraphics[width=0.5\textwidth, angle=0]{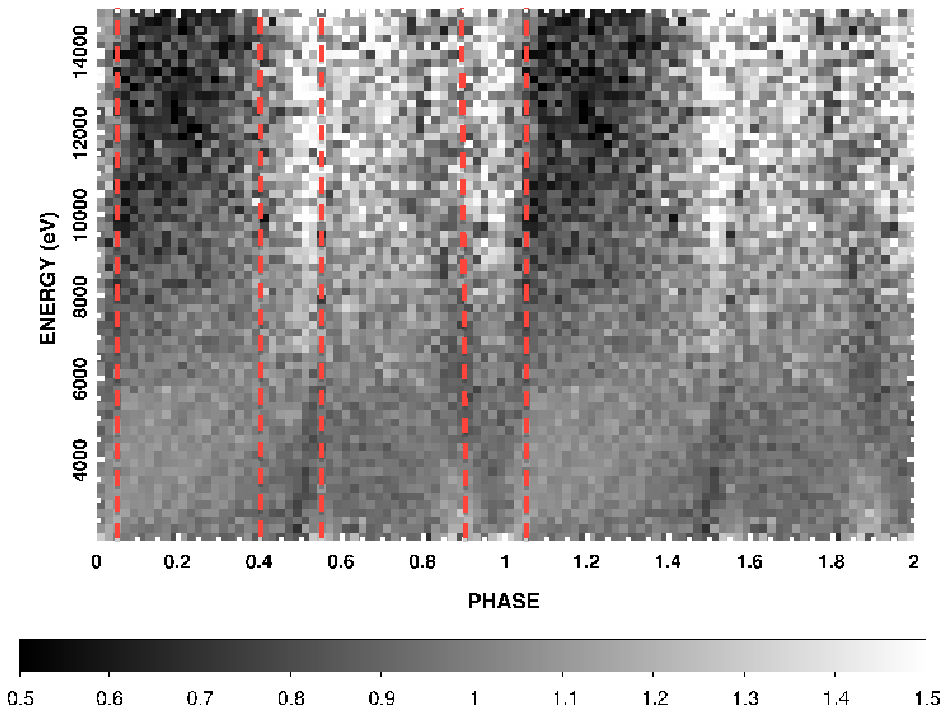}
  \includegraphics[width=0.5\textwidth, angle=0]{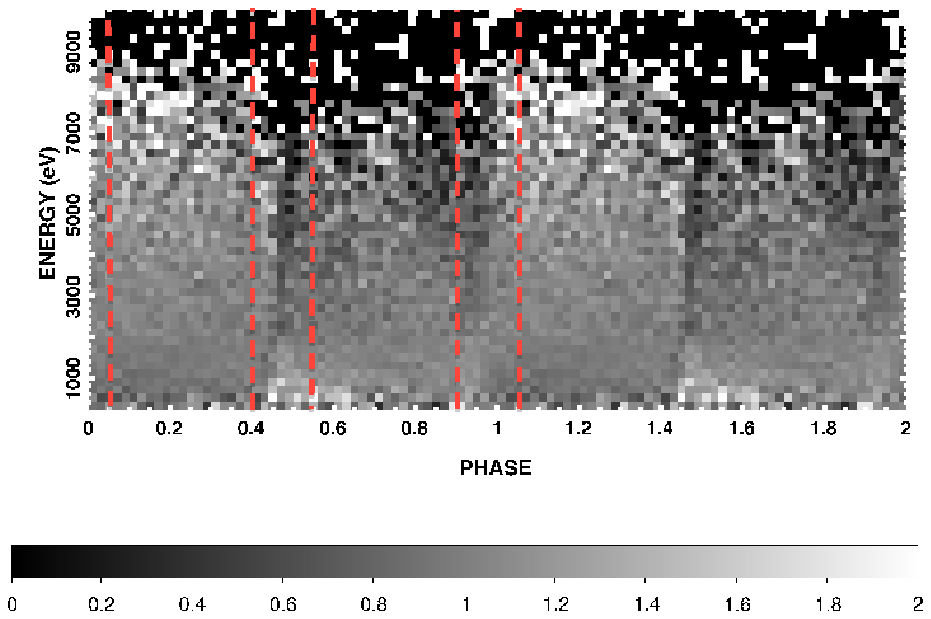}
  \includegraphics[width=0.5\textwidth, angle=0]{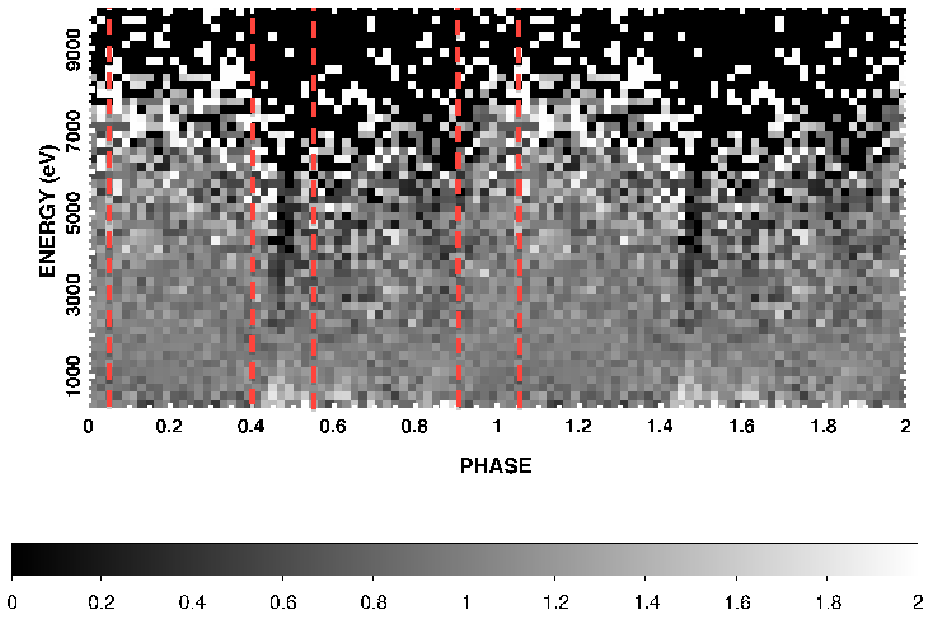}
  \caption{Normalized phase-energy images, divided by the phase-average spectrum 
    and the energy -integrated pulse profile, of \sgrswift\ from \rxte\ (PCA observations 
    from 2011 July 16 to 2011 July 20, {\it top panel}), \cxo\ (ACIS-S observation in 
    CC mode on 2011 July 27, {\it middle panel}) and \xmm\ (EPIC pn observation on 2011 
    September 23 {\it bottom panel}) data. The vertical dashed lines denotate the intervals 
    used in the PPS (see Fig. \ref{tratio}).
  }
  \label{NormPhaseEnergy}
\end{figure}

To better explore the complex variability with phase of the X-ray spectrum of \sgrswift, 
we created phase-energy images from the X-ray observations with the best counting 
statistics and enough spectral/time resolution, by binning the source counts into 
small phase and energy intervals (see Figs.~\ref{PhaseEnergy} and \ref{NormPhaseEnergy}). 
This kind of analysis has proven useful to identify possible narrow spectral features that strongly 
vary with phase, like the absorption-like feature discovered in \sgrlowb\ (\citealt{Tiengo13}; 
hereafter T13).

A dark straight line, slightly inclined to the right, is visible at phase $\sim$0.5 in 
the \rxte, \cxo\ and \xmm\ images (Figs.~\ref{PhaseEnergy} and \ref{NormPhaseEnergy}). 
In addition, the uppermost panel of Fig.~\ref{NormPhaseEnergy} obtained with RXTE data 
of energies $>$ 4 keV, displays a dark 'V'at phase $\sim$ 0.9 -- 1.1, similar to the one 
detected in the phase-energy images of \sgrlowb\ (T13). The same feature cannot be clearly 
detected in the \cxo\ and \xmm\ data likely due to poorer statistics at high energies. 

These dark tracks are rather narrow and almost vertical, suggesting the presence  
of an absorption-like line in the spectrum, whose energy rapidly varies with phase. 
As in T13, 50 phase-resolved spectra for each of the three datasets 
were extracted and analyzed with simple spectral models. Due to the relatively small 
number of counts in each phase interval, all the 
\xmm\ 
spectra were consistent with the model of the phase-averaged spectrum, 
simply rescaled by a flux normalization factor. On the other hand, the 
\rxte\ phase-resolved spectra displayed significant differences, which 
motivated a detailed analysis.

\begin{figure}
  \centering
  \includegraphics[width=0.5\textwidth, angle=0]{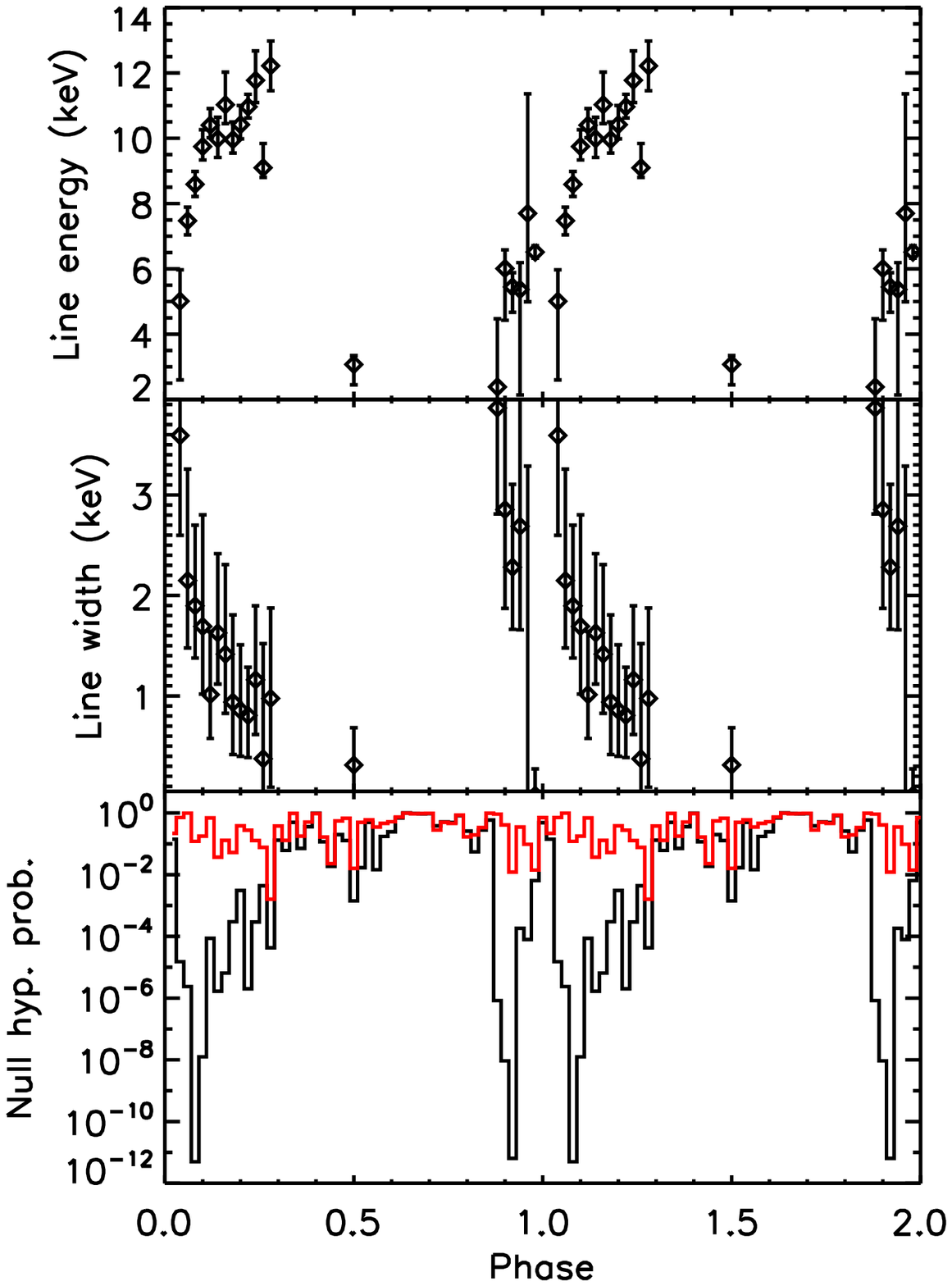}
  \caption{Results of the analysis of 50 phase-resolved spectra of \sgrswift\ during the 
    \rxte\ observations from 2011 July 16 to 2011 July 20. The energy and 
    width of the line is displayed in the upper panels for the phase intervals that were not 
    adequately fit (null hypothesis probability $<$0.01) by the model without absorption line.
    The bottom panel shows the null 
    hypothesis probabilities of the fits of each phase-resolved spectrum with the best fit 
    model of the spectrum extracted from the 0.6--0.8 phase interval (see text for details; 
    {\it black}) and after the addition of an absorption line ({\it red}). 
  }
  \label{LineVar}
\end{figure}

A template spectral model was derived by fitting the \rxte/PCA spectrum extracted 
from the 0.6--0.8 phase interval, where the phase-energy image shows no narrow-band 
spectral variability. For consistency with the previous analysis, we adopted a three 
blackbody model with photoelectric absorption fixed at 
$N_{\mathrm{H}} = 4.7 \times 10^{21}$ cm$^{-2}$ and one of the three blackbody components 
with $kT=150$ eV and $R_{\rm BB} = 10$ km (assuming a distance of 1.6 kpc).\footnote{Since 
the PCA spectra are analyzed only for $E>2.5$ keV, this soft blackbody component gives 
a negligible contribution to the spectral fit.} A good fit is obtained only by 
adding a Gaussian emission line with $E=6.47\pm0.06$ keV, $\sigma=0.3\pm0.1$ keV and flux 
of $(4\pm1)\times10^{-4}$ photons cm$^{-2}$ s$^{-1}$, possibly due to an incorrect 
reconstruction of the Galactic Ridge component in the PCA background model. 
In this case, the temperatures of the two variable blackbody 
components are: $kT_h=1.7\pm0.1$ keV and $kT_w=0.73\pm0.02$ keV. As can be seen in 
the lowermost panel of Fig.~\ref{LineVar}, this template model rescaled by an 
overall multiplicative factor (except for the Gaussian component, which, 
consistently with its background interpretation, is kept fixed in all spectra) 
does not adequately fit the spectra (null hypothesis probability $<$0.01) in 
the phase intervals 0.03--0.29, 0.49--0.51, and 0.87--0.99. Acceptable fits 
were obtained by adding a cyclotron 
absorption line feature (Makishima et al. 1990) to the model in these phase 
intervals. The resulting line centroids and widths are shown in Fig.~\ref{LineVar}.

\section{Results and Discussion}
\label{discussion}
We studied  the long-term evolution of the
magnetar \sgrswift, by following the source during its approach to
quiescence after  the outburst that led to its discovery on
2011 July 14. We performed a timing analysis by phase-connecting
observations covering about 500 days of the \sgrswift\ outburst
decay. The resulting values of the spin period and period
derivative are $P=8.437720019(7)$ s and 
$\dot{P}=1.34(1)\times10^{-13}$s\,s$^{-1}$, which, if interpreted in terms of
magneto-dipolar braking, lead to a value of the dipolar magnetic
field of $B \sim 3.4(1)\times10^{13}$ G at the equator of the star. This confirms that
\sgrswift\  is the magnetar with the second lowest dipolar
magnetic field detected so far.  By performing a timing
analysis, we found a significant second time derivative of the
period, {\it \"P}$=-5.1(2)\times10^{-21}$s\,s$^{-2}$, which is consistent with the upper limit reported
by R12 and with solution 2 of \citet{scholz12}; in \citet{Scholz14} a longer 
time span is split and fitted with an exponential glitch recovery for the first $\sim 60$ 
days and an additional $P$-$\dot{P}$ solution for the rest of their data set, 
obtaining a somewhat lower estimate of the dipolar magnetic field of 
$B \sim 1.4\times10^{13}$ G.

A closer look at the PPS results shows however that, apart from
the above mentioned common characteristics, the spectral evolution
of the source at phase intervals P1 and P3 is markedly
different from that observed at P2 and P4. The evolution of the
spectra at phase intervals P2 and P4 is very similar (see lower 
panels of Fig.s \ref{pps2.1} -- \ref{pps2.3}) in terms of
both measured values of BB radii and temperatures at each epoch, and in
terms of shrinking and cooling rates.
Instead, spectra observed at phase intervals P1 and P3 have
different initial values of the BB radii and temperatures and then
display different shrinking and cooling rates over the outburst
decay (see upper panels of Fig.s \ref{pps2.1} -- \ref{pps2.3}). 
This may suggest that, during the phase intervals P1 and
P3, we are observing two zones with physically distinct
properties, likely the two main surface zones that had been
heated during the outburst. Instead, during the phase intervals P2
nd P4,  we observe a transition, with one of the two heated zones
exiting and the second one entering into view. The fact that the
temperatures of P2 and P4 are consistently {\it in between} those
of P1 and P3  and that the shape of the observed pulse profile is
hardly compatible with a single hot spot, reinforces this two-zones 
interpretation. We therefore suggest the possibility that
the emission is coming from (at least) two zones of the neutron
star surface with different physical properties, the evolution of
which may be due to different mechanisms. The blackbody
radius $R_{BB}$ obtained from our spectral fits may not provide a
direct measure of the size $L$ of the emitting region on the star
surface. In fact, knowledge of the viewing geometry of the source (i.e.
magnetic and spin axis inclinations with respect to the line of
sight) and of the location of the emitting regions on the star
surface would be required to relate $L$ to $R_{BB}$. An evaluation of
the source geometry will not be attempted here (although this 
might be possible, see e.g. \citealt{Albano10}). 
We just note that, since these angles are not
expected to change during the outburst decay, the time evolution
of $R_{BB}$ mirrors that of $L$, that is to say that the rate of
change of the two quantities is the same.

The evolution observed during  the decay phase of an outburst is
usually interpreted in terms of either the cooling that follows 
deep crustal heating (\citealt{Perna11}) or untwisting of the star 
magnetosphere (\citealt{beloborodov09}). The overall behaviour  of 
\sgrswift\, i.e., the shrinking
of the heated zones at an almost constant temperature, is expected
in the framework of both models. For instance, in the untwisting
magnetosphere (UM) model,  the twist is initially implanted in a
current-carrying bundle of magnetic field lines which then
gradually shrinks. Since crustal heating is caused by back-flowing
currents in the j-bundle, a decrease in the size of the j-bundle
also implies a decrease in the heated surface area
(\citealt{beloborodov09}). We also note that the presence of a
negative $\ddot P$, as follows from our timing analysis, is in
agreement with the UM. In fact, immediately prior to the outburst,
when the magnetic field of the star is expected to be  highly
twisted, the star is subject to a large amount of spin-down torque
and therefore the spin-down rate is larger with respect
to its value in quiescence. If, later during the decay, the field
untwists, then the spin-down rate should return to the (secular)
pre-outburst value. Therefore in this phase a negative second
derivative of the spin period should be observed.

\begin{figure}
  \centering
  \includegraphics[width=0.35\textwidth, angle=270]{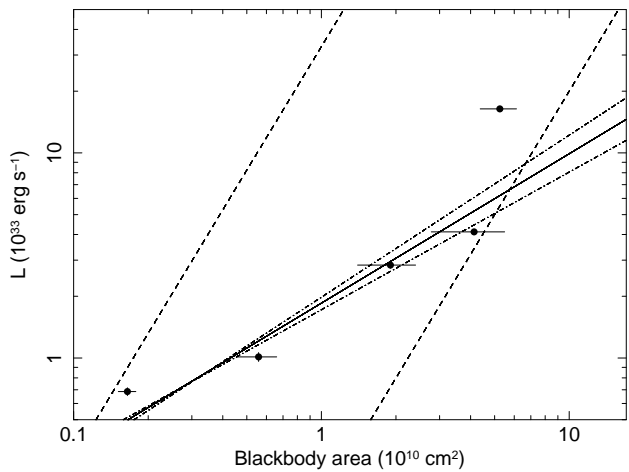}
  \caption[Luminosity versus emitting area of the P3 warm blackbody
    component]{Luminosity versus blackbody emitting area of the P3 warm
    component (See Fig. \ref{tratio} for reference).
    The dashed lines represent $L \propto A^{2}$, simple
    untwisting magnetosphere models, see \citet{beloborodov09}.
    The solid line is a power-law fit to the data, which yields
    $L \propto A^{0.7}$. The $3\sigma$ uncertainty of the fit parameters
    is represented by dot-dashed lines.}
  \label{L.vs.A3}
\end{figure}
\begin{figure}
  \centering
  \includegraphics[width=0.35\textwidth, angle=270]{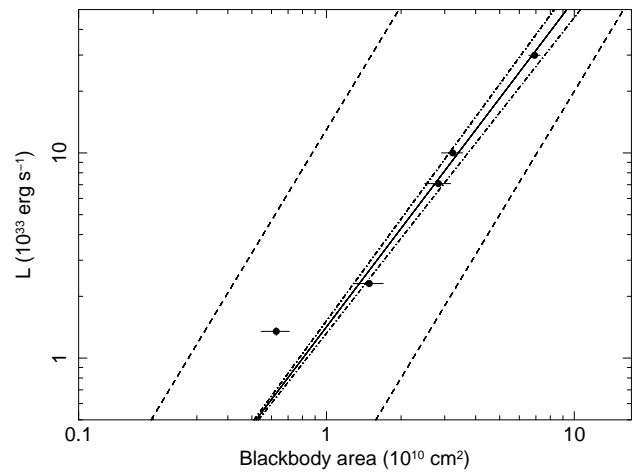}
  \caption[Luminosity versus emitting area of the P1 warm blackbody
    component]{Same as Fig. \ref{L.vs.A3}, for P1. The solid line is a
    power-law fit to the data, which yields $L \propto A^{1.6}$.
    The $3\sigma$ uncertainty of the fit parameters (corresponding
    to $L \propto A^{1.5}$ and $L \propto A^{1.7}$) is represented
    by dot-dashed lines.}
  \label{L.vs.A1}
\end{figure}

To further test the  possibility that a surface zone heated by the
j-bundle currents is visible at certain phase intervals, we
calculated the blackbody emission area and the luminosity
corresponding to the hot and warm components observed in the four
phase segments. A simple UM model predicts a relation between the
luminosity $L$ and the emitting area $A$ of the form $L \propto
A^2$ (see Beloborodov 2009). By assuming $L\propto A^n$, for the warm 
component we found a power 
law index of $n= 1.6, 1.0, 0.7, 0.9$ in the phase interval P1, P2, P3 and P4 
respectively. The correspondent index relative to the hot component are 
$n= 1.4, 2.3, 1.9, 1.0$. Therefore only some of the measured relations are 
close to the theoretical expectations. As an example, in Figs. 
\ref{L.vs.A3} and
\ref{L.vs.A1} we show the evolution of the luminosity with
respect to the emitting area as observed for the warm component
during phase intervals P3 and P1, respectively (the dashed lines
represent the relation expected theoretically). As it can be seen,
the evolution observed during P1 is compatible with being
related to an evolving current-carrying bundle, while the 
warm component observed during
P3, which evolves very differently ($L \propto A^{0.7}$), may be related 
to a different emission mechanism.
The luminosty/area relation detected at P3 is much flatter, and
seems to become steeper and steeper as the outburst decay
progresses. Our fitting model, which does not allow us to reconstruct the 
location of these components on the star surface, is however too 
simplicistic to derive more robust conclusions. 


An particular case is the evolution of the warm temperature
relative to the P3 phase interval (Fig. \ref{pps2.3}, upper
panel), which remains (within the 1-$\sigma$ errors) constant 
up to the second to last observation. Then in the last observation, 
the hot BB is not present anymore and 
the spectrum becomes consistent with a single free BB at $\sim 0.5$ keV
besides the fixed BB at 150 eV (and about 5km of radius).
This indicates that in this phase interval at later times 
only the warm region survives, possibly engulfing the hotter spot. 
Or, it can be interpreted in terms of the warm BB being slowly heated 
by the hotter region (hot BB), which dissipates (or shrinks out of the 
line of sight) as a result.


Fig. \ref{fluxdec} shows the flux evolution of the \src\ during
the outburst decay. If we assume the peak luminosity to be of a
few $10^{35}$ \ergs, as inferred by R12 based on magneto-thermal
evolutionary models, it implies a distance to the source of $\sim
2$ kpc. This is consistent with the distance to the Galactic
H\textsc{ii} region M17, described in Sec. \ref{specana}; and
provides further support for the \nh\ and distance values assumed
in this work.

We also reported the detection of a phase and energy-variable
spectral line (Figs. 10 and 11). As it can be seen in Fig. 10, a
dark V-shaped feature is visible in the RXTE data at phases
$\sim$ 0.9-1.1, followed by a dark area around phases 1.2-1.3. The
energy of the feature varies between $\sim$ 5 and $\sim$ 12 keV.
There is also another
feature at lower energy ($\sim$ 2 keV) at phase $\sim $ 0.5, which,
if significant, may be just a continuation of the main feature
(this is supported by the line width evolution but not by the line
energy evolution, see the two upper panels of Fig.11). A similar phase and energy
dependent feature has been detected so far only in two sources:
the low-B magnetar SGR~0418 (see T13) and the X-ray dim isolated
neutron star RX J0720.4-3125 (see \citealt{Borghese15}). 
s in these sources 
the feature may be due to 
proton-cyclotron resonant scattering of
X-ray photons emitted by the star surface onto charges flowing in
a small coronal magnetic loop (for alternative interpretations 
see the discussion in T13 and \citealt{Borghese15}). The energy 
variation of the line would be 
caused by magnetic field gradients along the coronal loop:  as the
neutron star rotates, photons directed toward us intercept
sections of the loop with different magnetic field intensities. In
the rest frame of the emitters, the proton cyclotron energy is
$E_{cp} = \hbar e B / m_{p}c \approx 0.63 B_{14}$ keV, where
$B_{14} = B/(10^{14} \text{G})$. Assuming that the line is
emitted near the surface of the star, its energy as measured by a
distant observer is significantly affected by gravitational
redshift. Thus, the magnetic field in the small corona loop can be
estimated as $B_{14} = (1+z) E_{obs}/0.63$ keV, where $E_{obs}$ is
the observed energy of the line, and $z = 2GM_{NS}/R_{NS}c^{2}
\approx 0.35$ (using a neutron star mass of $M_{NS} \approx
1.4M_{\astrosun}$) and a radius of $R_{NS} \approx 10$ km). In the
case of \sgrswift\ ($E_{obs}$ between 3 and 12 keV) the resulting
magnetic field in the coronal loop  is in the $ (6 - 25)
\times 10^{14}$ G range.

We note that the absorption line centroids, as inferred from the
RXTE PCA, are predominantly at energies close to the upper bound
of both the \xmm\ and \cxo\ spectral ranges, or even higher 
(Fig. 11, upper panel). Therefore,  due to the 
steep spectrum of \src\ and the rapid decrease of the
response of these instruments at energies above about 6 keV,
translating into small count statistics, this
feature becomes undetectable. In fact it would be virtually 
indistinguishable from a a modification of a broad spectral 
component (such as the hot  BB) when
observed in the \xmm\ or \cxo\ spectral range.

In the case of \sgrlowb\, instead, the line phase variability could be better
characterized because the line centroid energy extended down to $\sim 1$ keV, where
the \xmm\ effective area has its maximum. Moreover, the effect of
cyclotron scattering by currents in a magnetic loop is more pronounced 
when most of the X-ray radiation is produced by a single hot spot on
the magnetar surface, as in \sgrlowb\, rather than by two different
regions, as in \src.

The discovery of a second magnetar with a magnetic field in the
radio-pulsar range strengthens the idea that magnetars are much
more common than expected so far. Dormant magnetars may lurk among
unidentified, weak X-ray sources and reveal themselves
only when they enter an outburst phase. The detection of
phase-variable absorption lines in the spectra of both the
low-field magnetars discovered up to now is remarkable, the more
given that a similar feature has been observed only in the
thermally-emitting neutron star RX J0720.4-3125, which has a comparable
dipole field but has not exhibited any magnetar-like behaviour.
Whether this is a further proof of a (much suspected) link between
magnetars and the seven thermally-emitting isolated neutron stars
is still an open issue.

\section*{Acknowledgments} 
We acknowledge the use of public data from the \swift, \xmm\ and \cxo\ data archives.
PE acknowledges a Fulbright Research Scholar grant administered by the
U.S.--Italy Fulbright Commission and is grateful to the
Harvard--Smithsonian Center for Astrophysics for hosting him during
his Fulbright exchange.

\bibliography{bib}
\bibliographystyle{mn2e}

\end{document}